\documentclass{kapproc}

\usepackage{graphicx}

\normallatexbib

\begin{document}

\articletitle[Cytokinesis: an initial linear phase and a
multiplicity of endings]{Cytokinesis: the initial linear phase
crosses over to a multiplicity of non-linear endings}

\articlesubtitle{Biphasic cytokinesis and cooperative single cell
reproduction}

\author{David Biron}
\affil{Department of Physics of Complex Systems\\
Weizmann Institute of Science}

\author{Pazit Libros}
\affil{Department of  Biological Chemistry\\
Weizmann Institute of Science}

\author{Dror Sagi}
\affil{Department of Physics of Complex Systems\\
Weizmann Institute of Science}

\author{David Mirelman}
\affil{Department of  Biological Chemistry\\
Weizmann Institute of Science}

\author{Elisha Moses}
\affil{Department of Physics of Complex Systems\\
Weizmann Institute of Science}

\begin{abstract}
We investigate the final stage of cytokinesis in two types of amoeba,
pointing out the existence of \emph{biphasic} furrow contraction.
The first phase is characterized by a constant contraction rate,
is better studied, and seems universal to a large extent.
The second phase is more diverse.
In {\it Dictyostelium discoideum} the transition involves a change in
the rate of contraction, and occurs when the width of the cleavage furrow is
comparable to the height of the cell.
In {\it Entamoeba invadens} the contractile ring carries the cell
through the first phase, but cannot complete
the second stage of cytokinesis.
As a result, a cooperative mechanism has evolved in that organism,
where a neighboring amoeba performs directed motion towards the dividing cell,
and physically causes separation by means of extending a pseudopod.
We expand here on a previous report of this novel chemotactic signaling mechanism.
\end{abstract}


\section{Introduction}

Cytokinesis is the last step in cell division, during which the
physical separation of a mitotic cell into two daughter cells is
achieved. In the process a constriction appears in the cell
circumference, perpendicular to the mitotic spindle and at
mid-cell. An acto-myosin contractile ring attached to the inner
surface of the cell membrane deepens the cleavage furrow to
achieve daughter cell separation. In order to achieve a symmetric
partitioning this process must be highly controlled both spatially
and temporally.

Cytokinesis has been, and still is, a very active field of research.
Recently there have been several developments covering a wide range
of phenomenon implicated in the process.
The role of microtubules (e.g. the mitotic spindle) is better
understood, and it is now known that in many eukaryotes midzone
microtubule bundles are important for both spatial positioning of the contractile
ring and are continuously required for the progression of the cleavage furrow
\cite{Bray88,Wheatly96,Fishkind96,Eckley97,Zang97,Neujahr98}.

Signaling pathways for spatial and temporal regulating of furrow
assembly and ingression are continuously being discovered.
Rho-mediated signaling, for instance, is required for initiation
of cytokinesis in {\it Drosophila} \cite{Prokopenko99}, and for
recruitment of actin myosin to the furrow in {\it Xenopus} embryos
\cite{Drechsel97}.

An understanding of the roles of membrane anchoring, membrane
dynamics, and the precise role of myosin and other actin binding
proteins is also evolving. For example, the ability of myosin II
null {\it D. discoideum} to perform cytokinesis showed that
systems where myosin II is redundant indeed exist
\cite{DeLozanne87,Neujahr97,Uyeda00}. Reviews of these developing
topics can be found in \cite{Robinson00,Field99,Wolf99,Glotzer97}.

Finally, the mechanistic design principles of the actin
contractile ring have been of interest
\cite{Mabuchi86,Pollard90,Satterwhite92,Fishkind93,Pelham02}
although a comprehensive mechanistic model has still not emerged.
In summary, despite this impressive progress, much remains unclear
about the biochemistry, molecular pathways, regulation and
mechanics of cytokinesis.

A genetically tractable model system which has been extensively studied
is the cellular slime mould, {\it Dictyostelium Discoideum}
\cite{Weber01,Neujahr97,Neujahr97b,Gerisch00}.
%Although the degree universality of the conclusions drawn
%from this model system is not known,
This system provides a clear example of two distinct types of cell
cycle coupled cytokinesis, namely ``cytokinesis-A'' and
``cytokinesis-B'' \cite{Zang97,Uyeda00b,Nagasaki01,Nagasaki02}.
Cytokinesis-A is characterized by an adhesion independent, myosin
II driven constriction of the cleavage furrow (e.g. as observed in
cells in suspension), while cytokinesys-B, observed mainly in
myosin II null {\it Dictyostelium} mutations, exhibits adhesion
dependent, myosin II independent separation furrow constriction.
The relative contribution of each of the two mechanisms in wild
type adherent cells has not been determined.

Other mutations of {\it Dictyostelium} cells have been utilized
to study different aspects of cytokinesis.
The small GTPase RacE, for example, was found to be crucial for furrow progression
throughout cytokinesis in suspension \cite{Gerald98}.
GAPA, a RasGTPase-activating protein encoded by the {\it gap}A gene,
was reported to be specifically involved in the completion of cytokinesis
\cite{Adachi97}. Similarly, depletion of Dynamin A by gene-targeting
techniques obstructs the completion of {\it Dictyostelium} cytokinesis \cite{Wienke99}.

The dramatic changes of cell shape during division are indicative of the mechanical
forces exerted at the cleavage furrow.
There are various measurements of a constant rate contraction phase
which lasts until the width of the waist connecting the daughter
cells is $\sim 1/10$ of the initial cell diameter \cite{Mabuchi94,Schroeder72}.
The steady contraction ensures smooth conditions in the process,
but as the connection between
daughter cells is about to vanish a singularity is produced,
making control non trivial.

At this stage the contractile ring is reported to disintegrate \cite{Schroeder72},
and the effect of physical properties of the connecting waist
(e.g. elasticity and viscosity) on the contraction becomes increasingly dominant.
The contraction therefore enters a second, non linear phase,
which has remained particularly enigmatic \cite{Robinson00}.
While the first part of our paper deals with the biphasic structure of
cytokinesis, in the second part we proceed to
show what happens in the second phase of the particular case of {\it E. invadens}.

{\it Entamoeba invadens} is a highly motile amoebic parasite originating from reptilian
intestines, used as a model for the study of encystation by the human pathogen
{\it Entamoeba histolytica} \cite{Wang03}.
As we have demonstrated previously \cite{Biron01}
the {\it E. invadens} daughter cells can complete cytokinesis by becoming motile,
pulling apart from each other and stretching the connecting tether until it breaks.
Alternatively, they can cooperate by employing a chemotactic
mechanism, i.e. a neighboring cell follows the concentration gradient
of a chemical secreted by the dividing cell and physically separates
it into two viable daughter cells.

\section{Materials and Methods}
\label{Sec::MandM}
\subsection{Cells}

{\it E. invadens} was grown at $25-27^{o}C$ in air tight $42  \
ml$ plastic flasks containing TYI-S33 medium ($870  \ ml$ nanopure
water, $20  \ g$ trypticase, $10  \ g$ yeast extract, $10  \ g$
glucose, $2  \ g$ NaCl, $1  \ g$ K$_{2}$HPO$_{4}$, $0.6  \ g$
KH$_{2}$PO$_{4}$, $1  \ g$ cystein hydrochloride, $0.2  \ g$
ascorbic acid, $22.8 mg$ ferric ammonium citrate, $130  \ ml$ heat
inactivated bovine calf serum, $30  \ ml$ Diamond vitamin tween
$80$ solution $40X$ from JRH Biosciences). Observing the cells at
$100X$ magnification was enabled by making a hole on the bottom
wide face of the flask and sealing it with a thin cover slip glued
with paraffin wax.

{\it D. discoideum} was grown at $24^{o}C$ in petri dishes
containing HL5 medium ($14.3  \ g$ peptone, $7.15  \ g$ yeast extract,
$18  \ g$ maltose, $0.64  \ g$ Na$_2$HPO$_4$ and $0.49  \ g$ KH$_2$PO$_4$
per liter, pH $7.0$). Using $40X$ magnification required
replacing part of the bottom face of the dish with a cover slip, as
described above.
While motility stops concurrently with division, the
observation was initially obscured by the detachment of the connecting section
between the two daughter cells from the substrate, and a subsequent loss of
focus.
Hence, the cells where overlayed with a $2 \%$
agarose sheet following the agar-overlay technique
\cite{Yumura85,Fukui91}.
This allows imaging of the cleavage furrow in
the amoeba during the full division time, while allowing the amoeba to live
normally between divisions.

\subsection{Chemotaxis assay}

A $3-4  \ cm^{2}$ hole was made on the top wide face of the flask
(i.e. above the face where the cells were plated). This hole was
fitted with a glass cover that was secured in place by paraffin
wax prior to plating of cells. Since {\it E. invadens} are
%microaerophillic,
aerotolerant anaerobes,
it was essential to preserve their reduced oxygen environment
throughout each experiment.
Removing the glass cover and then quickly filling the hole with
mineral oil shortly before each experiment started
enabled us to easily insert glass pipettes into the flask
while maintaining a low rate of air diffusion into the medium.
A glass pipette, $10-15  \ \mu m$ in diameter at the tip,
was used for locally aspirating and discharging the assayed medium
as described in \cite{Biron01}.  The pipette was held in a fixed
position, whereas the flask was placed on a moving stage mounted
on a Zeiss Axiovert 135TV inverted microscope.
The position of the stage in the horizontal plane was manually
controlled by a pair of motorized actuators.
All experiments were recorded using a CCD camera at video rate.

\subsection{Collection of medium for chemotaxis assays}

We collected small volumes ($\sim 10 \ pl$) of attractant
containing liquid by aspirating medium from the vicinity of the
furrow of a dividing amoeba, using a glass micro-pipette with a
tip diameter of $5-10 \mu m$. We also took $1 - 1.5 \ ml$ samples
of fluid from the bottom of a $4-$day old flask with an (almost)
confluent culture. At this stage the culture is peaking in its
division cycle, with over $10^{6}$ divisions per flask per day.
This fluid was either directly assayed for attractive activity or
subjected to size analysis (e.g. Millipore centricon filtering and
SDS-PAGE analysis). The active fraction was further chemically
treated as described below.

\section{Results}
\subsection{Cytokinesis in wild type {\it D. discoideum} is
biphasic}

The accumulated data from $15$ wild type cell divisions in {\it D.
discoideum} is shown in Fig.~1. We show only cases where no motile
activity interfered with division, and discuss motile separation
below. We plot the behavior of the furrow width ($D_m$) as a
function of $\tau = (t_{f}-t)$, where $t_{f}$ is the time of
complete separation. A clear transition is observed at the
critical time $t_{c} \sim t_{f}-70$ seconds.

\begin{figure}[ht]
\label{FIG1::grf}
\begin{picture}(12,155)(0,0)
\includegraphics[width=120mm]{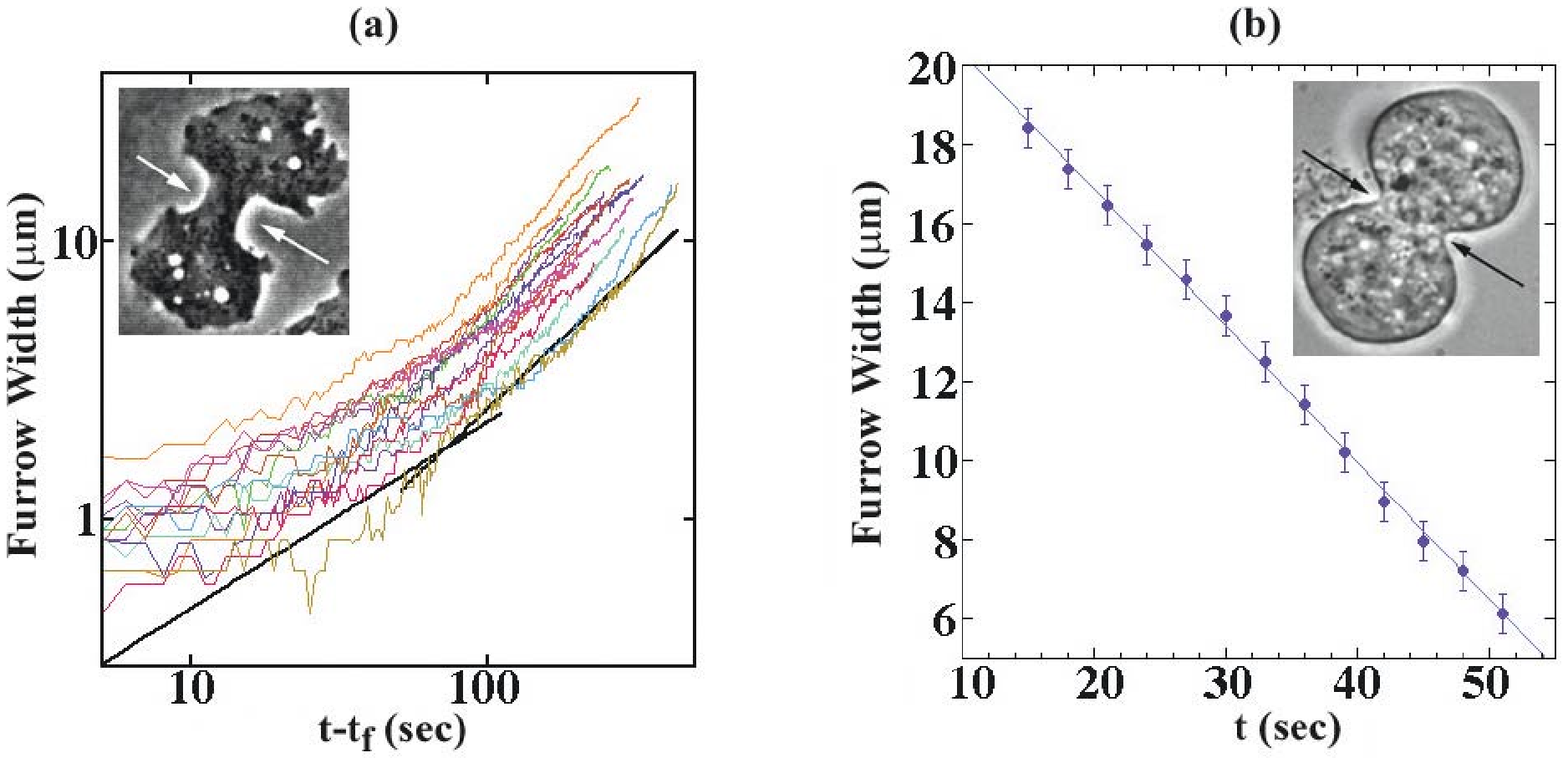}
\end{picture}
%\sidebyside
         {\letteredcaption{a}{A log-log plot of \emph{D. discoideum} furrow
             width, $D_m$, (under an agar overlay). At time $t_{f}$
             separation of the two daughter cells is completed.}}
         {\letteredcaption{b}{\emph{E. invadens} furrow width, $D_m$,
             as a function of time. At time $t = 0$ furrowing
             begins.}}
\end{figure}
\addtocounter{figure}{1}

 The furrow width exhibits two
regimes which can be characterized by two different values of the
exponent $\alpha$ in the power law $D_{m} \propto \tau^{\alpha}$.
Separating the two time regimes, fitting them to a power law
behavior and then averaging exponents over the $15$ divisions
gives $\alpha_{1}=1.16\pm 0.2$ for the initial stage, and
$\alpha_{2} = 0.65 \pm 0.1$ for the final stage of separation.
Previous measurements of the cleavage furrow as a function of time
have been carried out, for example on dividing eggs of sea urchins
\cite{Mabuchi94}, and the width of the cleavage furrow was shown
to decrease linearly ($\alpha \approx 1$) for $D_m$ was $10\% -
20\% $ of the original cell diameter. However, the final stage was
not previously presented, e.g. the data presented in
\cite{Mabuchi94} does not reach full separation.

A different type of division, in which the two daughter cells regained motility
before complete separation in a fashion resembling cytokinesis-B \cite{Robinson00,Neujahr97},
was recorded for $\sim 20 \%$ of the {\it D. discoideum} divisions
under an agarose sheet.
At the final stages of these cases, the daughter cells would move in opposite directions,
parallel to the connecting tether, stretching it by applying tension forces along its
long axis until it ripped.
This process  was typically slower than cytokinesis-A dominated separation.
We conclude that {\it D. discoideum} furrowing
exhibits two regimes: a linear regime and a second which can be either
non-motile and accelerating, or motile and prolonged.

\subsection{Cytokinesis in {\it E. invadens}}

We have measured the dynamics of cytokinesis of {\it E. invadens}.
There also we observe a linear initial stage, but contrary to the
case of {\it D. discoideum} -- we never saw a non-motile second
phase in {\it invadens}. Instead, a transition to motile
separation always occurred at the end of the linear stage. Fig.~1
shows the furrow dynamics up to the time when $D_{m}(t) \approx 6
\  \mu m \sim 0.25 D_{m}(0)$. The conclusion of the process could
not be measured since {\it E. invadens} did not enter the same
non-linear part of cytokinesis. Instead, the contraction of the
ring halted, sometimes for several minutes.

As we reported previously \cite{Biron01}, we were able to discern
three alternative processes that continued beyond the arrest in
the cleavage process. In all cases cytokinesis-A was abandoned
after the pause. In a majority of the cases cytokinesis was either
reversed, or resumed by an application of traction forces by the
two daughter cells, actively advancing in opposite directions.
This resembles the phenotypes of several mutations known to cause
cytokinesis defects in other cells, as we discuss in the next
section.

\subsection{Cooperative behavior -- ``Midwife'' statistics}

We have recorded $n=106$ {\it invadens} divisions from four
different samples, all at a density of about $3\cdot 10^{4}$
cells/cm$^2$ and without an agar overlay. The results are
summarized in Fig.~2a. During the linear phase there is no motile
activity in any of the daughter cells. As the second stage is
approached, the connective cylinder between the two daughter cells
detaches from the substrate and cleavage stalls. At this point
both daughter cells resume motility and begin to move away from
one another, pulling at the tether that keeps them bound to each
other.

\begin{figure}[ht]
\label{FIG2::grf}
\begin{picture}(12,150)(0,0)
\includegraphics[width=120mm]{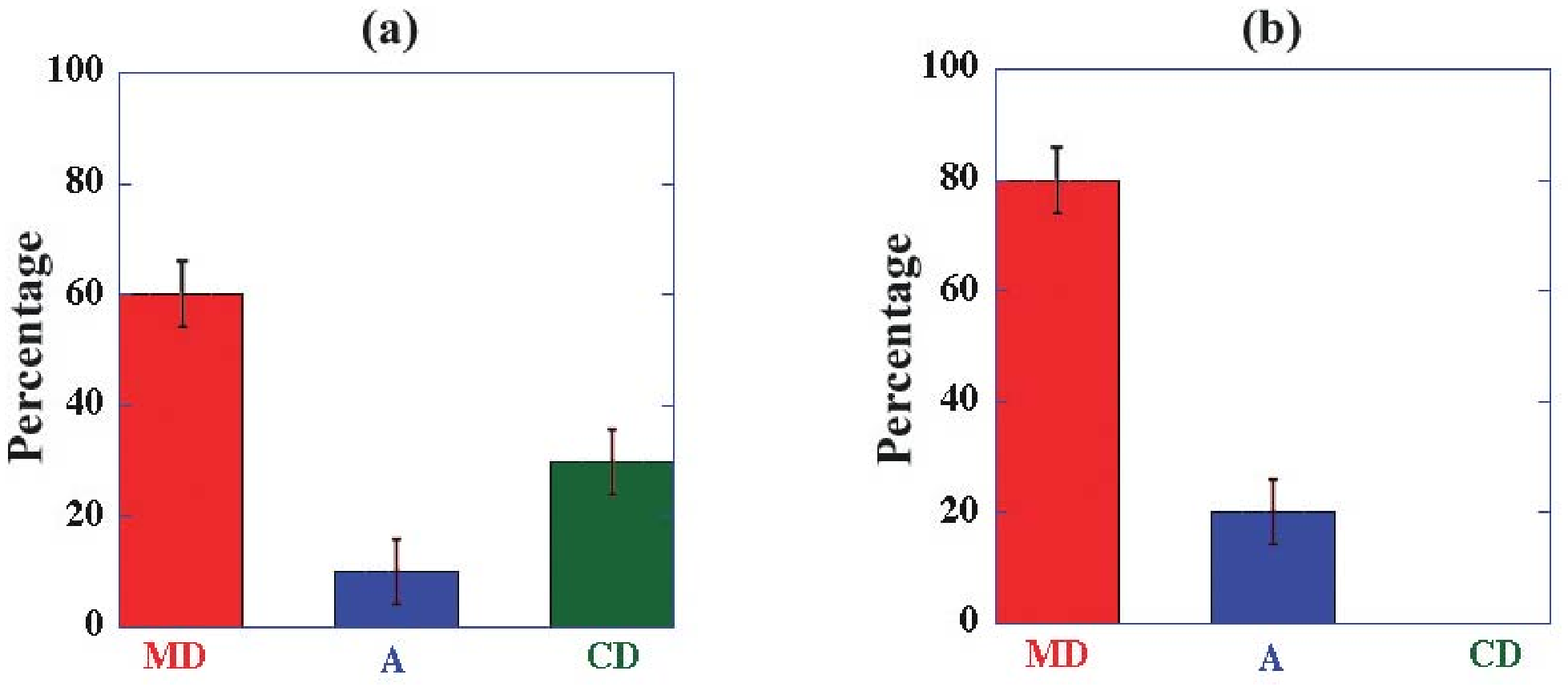}
\end{picture}
\letteredcaption{a}{
          The distribution of the three observed terminations of
          \emph{invadens} cytokinesis. MD -- motile divisions;
          A -- abortions; CD -- cooperative divisions (``midwives'').
          The distribution was measured at a cell density of
          $3\cdot 10^{4}$ cells/cm$^2$.}
\letteredcaption{b}{
          The same distribution measured at a cell density of
          $5\cdot 10^{3}$ cells/cm$^2$.
          Midwives were not detected at the lower cell density.}
\end{figure}

In $63$ cases ($60\pm 6 \%$) motile separation of the amoeba was
observed: the tension on the tether increased as they pulled and
it grew longer and narrower until it was cut. Another $11$ ($10\pm
2 \%$) cases ended in ``abortions'', where cytokinesis stopped and
the amoeba continued to live as a multi-nucleated cell. The cells
may then resume division and we have seen division into three or
four viable daughter cells. These types of behavior have also been
previously observed in the myosin II null mutants of {\it D.
discoideum} \cite{Neujahr97}.
A plausible interpretation is that in {\it E. invadens} the %*******
function of the contractile ring is incomplete.             %*******

Surprisingly, in the remaining $32$ ($30\pm 6\%$) cases a
neighboring amoeba intervened in the process of division. This
"midwife", as we call it, can travel a long distance (we have
measured directed motion of up to $200  \  \mu m$), usually in a
straight trajectory to the dividing amoeba. It aligns itself
parallel to the dividing amoeba, sends a pseudopod in-between the
separating amoebae and then severs the connection by moving itself
forcibly into the gap and pushing them apart \cite{Biron01}. This
process is depicted in Fig.~3.

\begin{figure}[ht]
\label{FIG3::grf}
\begin{picture}(12,200)(0,0)
\includegraphics[width=120mm]{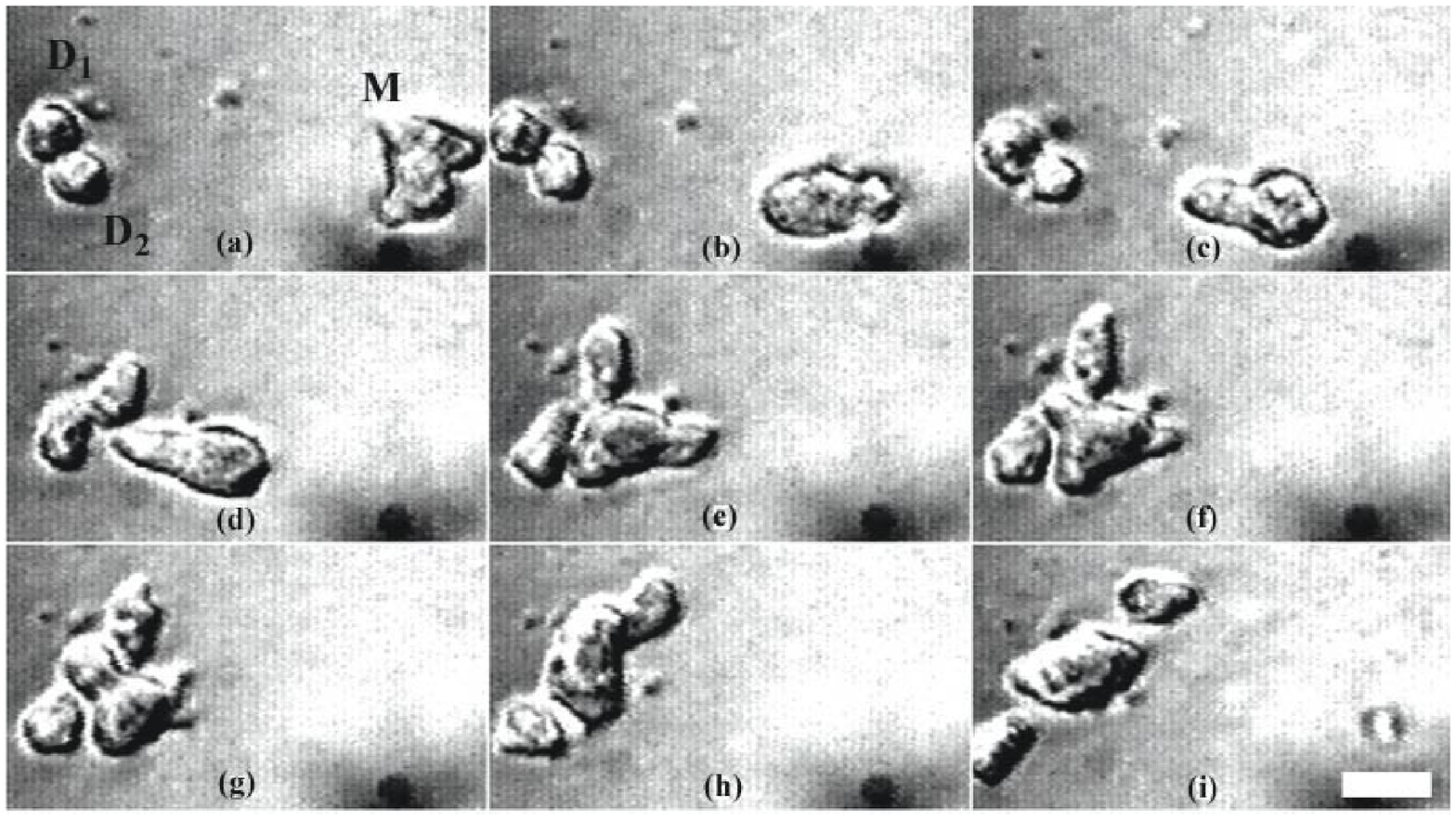}
\end{picture}
\caption{    A ``midwife'' (labeled ``M'') advancing towards a
             dividing cell and cutting through the tether connecting the
             two daughter cells (labeled ``D$_{1}$'' and ``D$_{2}$'') at the
             final stage of \emph{invadens} cytokinesis.
             Frames were taken at $t = 0$, $50$, $74$,
             $123$, $136$, $142$, $146$, $162$ and  $184$ seconds. %\\
             Scale bar is $10  \  \mu m$.}
\end{figure}

\subsection{The "midwife" is chemotactic}

In order to determine whether chemotaxis accounts for the long
directed trajectory that the midwife travels we aspirated $\sim 10
\ pl$ of medium from around a dividing amoeba using a
micro-pipette (Fig.~4). This medium was then released in a
controlled manner near a distant amoeba. In over $50\%$ of the
cases a chemotactic response was found, and the amoeba followed
the pipette as we moved it away for distances of up to a few
hundred micro-meters.

\begin{figure}[ht]
\label{FIG4::grf}
\begin{picture}(12,75)(0,0)
\includegraphics[width=120mm]{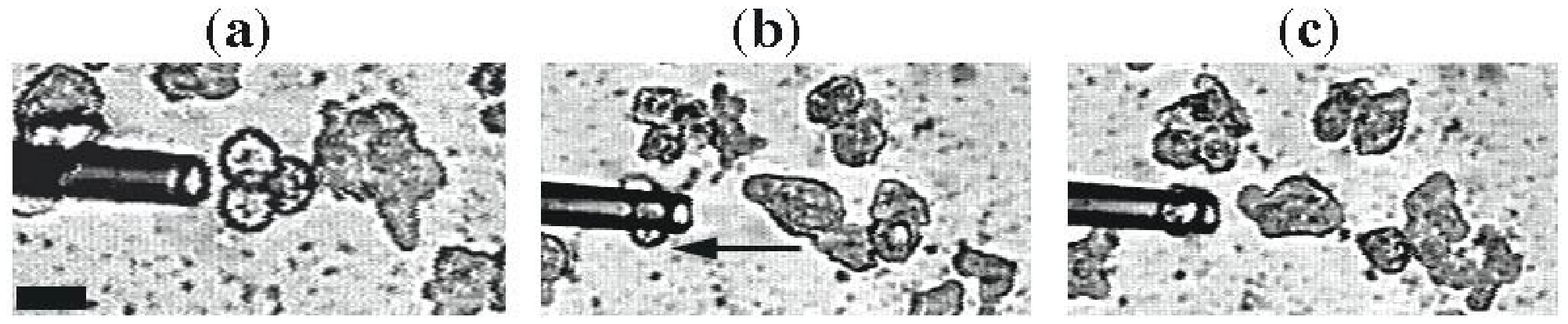}
\end{picture}
          {\letteredcaption{a}{Aspirating $\sim 10 pl$ medium from the vicinity of a
             dividing \emph{E. invadens}. Using a micro-pipette,
             the fluid was then released near far-off amoebae.
             As cells approached the secreting pipette, it was
             pulled away from them.}}
         {\letteredcaption{b}{An amoeba following the secreting pipette.
             The arrow shows the direction of movement.
             This cells followed the retracting pipette for
             $\sim 200  \  \mu m$.}}
         {\letteredcaption{c}{The same cell closer to the tip
             after  $20 sec$ in which the pipette was held in place.
             Scale bar is $10  \  \mu m$.}}
\end{figure}

Movies showing these measurements appear at \cite{NatureWebSite}.
In the cases when chemotaxis did not work, we could invariably
point out factors of the experiment that could have made it faulty
(e.g. the pipette was too thin) but we could not exclude the
possibility that only part of the dividing cell population
secretes an attractant, or that midwives react only in a specific
stage of the cell cycle. Controls such as aspirating near
non-dividing amoeba or using regular medium gave no response.

To obtain larger volumes of medium containing attractant, we
relied on chemoattractants being long lasting (unless a specific
enzyme is targeted to degrade them) \cite{Eisenbach03}, and took
$1 - 1.5 ml$ samples of medium from a flask containing
confluent $4$-day old culture, as described in Sec.~2. %\ref{sec::MandM}
We assayed these samples, and got a strong chemotactic response
(Fig.~5). A control experiment using $4$-day old medium from the
same batch, in which no cells were grown, did not elicit any
response.

\begin{figure}[ht]
\label{FIG5::grf}
\begin{picture}(12,90)(0,0)
\includegraphics[width=120mm]{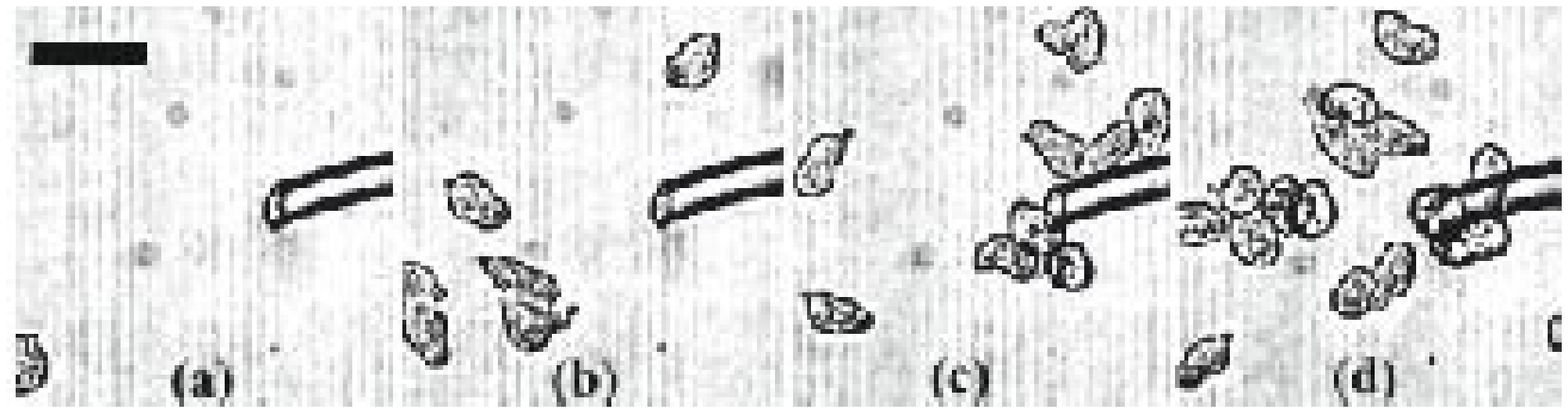}
\end{picture}
\caption{  \emph{E. invadens} cells in fresh medium
attracted to a
           stationary micro pipette secreting medium originating from a
           $4-$day old almost confluent culture. Scale bar is
           $50  \  \mu m$. $t = 0$, $375$, $750$ and $1290$ $sec$.}
\end{figure}

The assay for chemotaxis was quantified by comparing the
velocities of {\it E. invadens} cells that were suspected to be
stimulated  by the ``midwife'' calling signal and non-stimulated
cells. Fig.~6 depicts a set of such measurements.

\begin{figure}[ht]
\label{FIG6::grf}
\begin{picture}(12,170)(0,0)
\includegraphics[width=120mm]{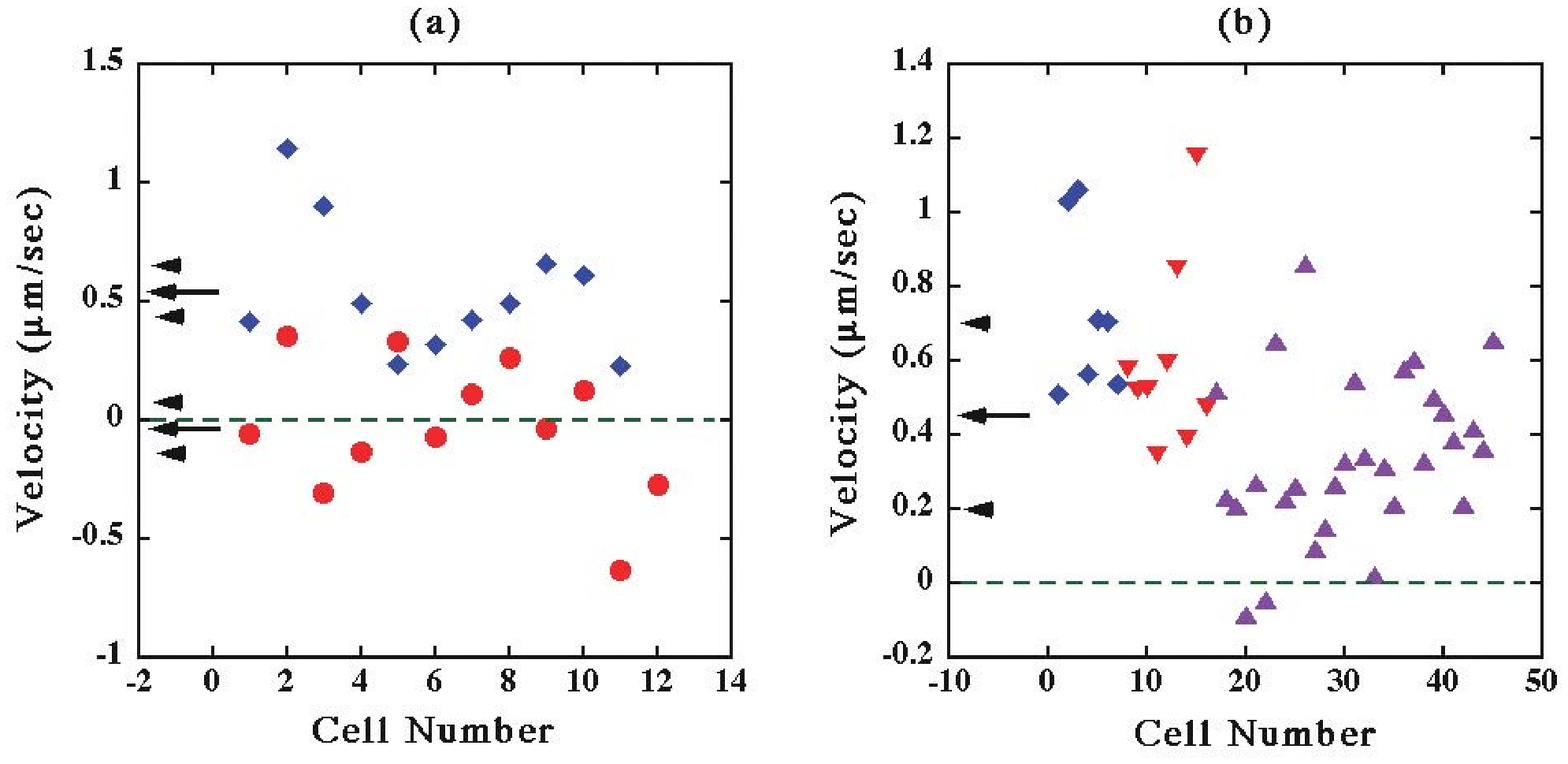}
\end{picture}
%\sidebyside
         {\letteredcaption{a}{
             Diamonds denote velocities of \emph{E. invadens} cells stimulated by a
             retracting pipette containing medium from the vicinity of a dividing
             amoeba ($v_{stim} = 0.5 \pm 0.05  \  \mu m/sec$); Circles denote
             control experiments done in the same fashion but with
             a pipette filled with plain medium, i.e. cells
             were not stimulated
             ($v_{non} = 0 \pm 0.05  \  \mu m /sec $).
             The arrows show the average velovity $\pm$ one standard
             deviation.}}
         {\letteredcaption{b}{Velocities of \emph{E. invadens} cells advancing towards a
             pipette containing medium extracted from a confluent culture flask.
             Diamonds denote horizontal velocities towards a retracting
             pipette. Triangles denote radial
             velocities towards a stationary pipette.
             The arrows show the average velovity $\pm$ one standard
             deviation.}}
\end{figure}

We have found the average stimulated cell
velocity to be $v_{stim} = 0.5 \pm 0.05  \  \mu m /sec $, while the
non-stimulated cells were
randomly diffusing about at $v_{non} = 0 \pm 0.05  \  \mu m /sec $. The
significant difference (an order of magnitude above the noise) enabled us to
use cell velocities to distinguish directed from random motion. These
results compared well to the velocity of directed motion towards a source of
high glucose concentration which was $\sim 1  \  \mu m /sec $ (the exact value
depended on various parameters e.g. the glucose concentration, the geometry of
the pipette and the initial velocity of the flow out of the pipette).

\subsection{Density dependence of cooperativity}

There is a strong dependence of the statistical distribution of
divisions on factors such as cell density, medium quality etc. In
a control experiment we reduced the cell density from $3\cdot
10^{4}$ cells/cm$^2$ to $5\cdot 10^{3}$ cells/cm$^2$ and measured
the distribution of cytokinesis terminations (Fig.~2). Such a
reduction in cell density increases the average distance between
cells $\sim 2.5$-fold, so that it is $\sim 150 \mu m$. As a
result, two factors hinder the chemotactic response. When the
distance of diffusion exceeds a few hundreds of micro-meters the
chemoattractant reaches its target highly diluted. In addition, as
we will demonstrate, the time it takes a neighboring cell to
respond to the signal becomes too large.

Our preliminary data shows the chemoattractant to have a molecular
weight of $50 \pm 10 kDa$. A globular molecule of this size will
diffuse a distance of $150 \mu m$ in an average time of $t_{D}
\approx 2 min$ in water at $25^{o}C$. A midwife moving at a
velocity of $\sim 0.5 \mu m/s$ would travel this distance in
$t_{T} \approx 5 min$. Therefore the time it would take for a
signal to get to a neighbor cell, and for the cell to respond and
approach would be $t_{D}+t_{T} \geq \Delta t$, where $\Delta t
\sim 5-10 min$ is the typical time interval between the release of
the signal and the completion (in some way) of cytokinesis. This
increases the probability that the dividing cell will either
complete its separation or abort the process before help arrives.

When the midwife can no longer come to aid separation, the cases
which would have concluded in cooperation at higher densities are
divided between successful motile separations and abortions. If we
consider only the cases in which a ``midwife'' did not intervene,
we can see in Fig.~2, that at the high cell density approximately
one in every seven unassisted attempts to divide ends up in an
``abortion''. At the low cell density, however, as many as one in
every five unassisted attempts to divide fails i.e. cooperation
increases the ratio of successful motile (independent) separations
to abortions.

\section{Discussion}
\subsection{Cytokinesis in {\it D. discoideum}}

We have measured a transition in the scaling of {\it D.
discoideum} furrow contraction dynamics. After a short transient
the furrow contracts (approximately) linearly until its width is
of the order of the cell height. It then speeds up ($\alpha = 0.65
\pm 0.1$) towards the end.

Comparing our measurements of wild type cells to the furrow
dynamics of mutant {\it D. discoideum} suggests that biphasic
dynamics, i.e. a linear contraction phase throughout $80-90\%$ of
the furrow progression, followed by a short non-linear phase is a
signature of normal cytokinesis-A in adherent cells. In contrast,
data of furrow dynamics of adherent myosin II null cells (i.e.
cytokinesis-B), published in \cite{Zang97}, demonstrates the lack
of a linear contraction phase. In fact, the data published does
not exhibit a transition between two dynamic regimes, but can be
fitted by a single power law, $D_{m} \propto \tau^{\alpha}$ with
an exponent of $\alpha = 0.55 \pm 0.05$. For comparison, Zang et
al. present the furrow dynamics of $\Delta$BLCBS-myosin cells,
which are assumed to overexpress myosin. The data shows a linear
phase ($\alpha \approx 1$) which is even longer than in typical
wild type cells, ranging from a furrow width of $D_{m} \approx 11
\mu m$ to $D_{m} \approx 1 \mu m$.

Another mutation of interest is dynamin A null {\it D. discoideum},
i. e. {\it dym}A$^{-}$ cells \cite{Wienke99}.
These cells are able to grow in suspension, form
a functional contractile ring during cell division, and pass through
all stages of cytokinesis up to the point where only a thin cytoplasmatic
bridge is connecting the two daughter cells. The final severing of
this bridge is inhibited by lack of dynamin. The published data on
{\it dym}A$^{-}$ cells shows a biphasic dynamics, with an initial
linear phase and a nonlinear ($\alpha < 1$) second phase.

Finally, the protein GAPA is specifically involved in the
completion of cytokinesis \cite{Adachi97}.
Again, {\it D. discoideum} cells lacking GAPA  ({\it gap}A$^{-}$)
are reported to exhibit normal cytokinesis, including myosin
II accumulation in the cleavage furrow, until the step in
which the daughter cells should be severed.
In light of the cases reported above we can conclude that the presence of
a biphasic furrow dynamics qualifies as a quantitative measure of ``normal''
cytokinesis-A.

\subsection{Cytokinesis in {\it E. invadens}}

The linear contraction phase was also measured in {\it E.
invadens}. In this cell, however, the second phase and the
conclusion of successful cytokinesis is carried out utilizing
motility. In $\sim 30 \%$ of the cases the separation mechanism
involves the novel phenomenon of cooperation with a neighbor (non
dividing) ``midwife'', recruited by a chemotactic signal.

The arrest in cytokinesis of {\it E. invadens} at the end of the linear phase
resembles the phenotype of mutations known to obstruct cytokinesis in other cells.
It is therefore worthwhile to compare it with other organisms, and
specifically with cytokinesis deficient mutants.
Following furrow localization and the beginning of
constriction, such mutations commonly display an arrest in
ingression, and then either reverse cytokinesis to form a
multinucleated cell or abandon cytokinesis-A and use motility (in
opposing directions) to complete the separation between the daughter
cells.

For example, adherent {\it gap}A$^{-}$ cells (on a glass
substrate) initiated cytokinesis similarly to wild type cells until the
daughter cells were connected by a thin cytoplasmatic bridge
\cite{Adachi97}.
At this stage the bridge remained intact for a long time,
and in $\sim 50 \%$ of the cases cytokinesis was finally reversed.
The other cases completed separation using traction forces,
reminiscent of what happens in {\it E. invadens}.

Another similar phenotype consists of {\it D. discoideum} cells lacking racE.
These mutant cells could initiate furrowing but then failed to divide in
suspension \cite{Gerald98}.
A third example is the cells of {\it Caenorhabditis elegans} embryos deprived of
ZEN-4 (a homologue of the MKLP1 subfamily of kinesin motor proteins capable of
bundling antiparallel microtubules), which were shown to initiate the cytokinetic
furrow at the normal time and place. However, their furrow propagation
halts prematurely, and after a short pause it retracts,
ultimately producing a single multinucleated cell.

In contrast, when a checkpoint control is affected by a mutation in budding yeast
the phenotype is different. Act5 is a yeast actin related protein which is necessary
for dynein function, and hence is important for spindle orientation \cite{Muhua94,Muhua98}.
Mutant cells often misalign the mitotic spindle, slowing its entrance to the neck between
mother cell and budding daughter cell. This delays cytokinesis at a checkpoint
control. After the delay, if the spindle enters the neck, cytokinesis
is completed normally.

Following our report of the {\it E. invadens} ``midwife'',
similar cooperative behavior has been observed in other cells.
Mechanical intervention in cytokinesis in {\it D. discoideum}
was recently reported and is currently studied \cite{Insall01,Nagasaki02}.
In addition, fibroblast cells on a smooth surface were observed
cooperating in a similar fashion \cite{FrenchPrivateComm}.
To our knowledge, there have been no reported cases of similar ``midwiving''
behavior in a mammalian extracellular matrix. It would be extremely interesting
if this was not restricted to highly motile cells, but could also
occur in neighboring cells in tissue.

%\begin{acknowledgements}
\section*{Acknowledgements}
This investigation was supported in part by the Binational Science
Foundation under grant no. $2000298$, by the Clore Center for
Biological Physics, by the EU Research Training Network PHYNECS
and by a grant from the Center for Emerging Diseases, Jerusalem.
We thank Shoshana Ravid for providing us with the {\it D.
discoideum} and for advice on their culture and observation.
%\end{acknowledgements}

%\bibliographystyle{kapalike}
%\chapbblname{chapbib}
%\chapbibliography{T}

\bibliographystyle{kapalike}
%\chapbblname{chapbib}
%\begin{chapthebibliography}
%\input MidwifeIIProc

%\end{chapthebibliography}

\end{document}